\documentclass[10pt]{forma}

\usepackage{graphicx} 

\def\be{\begin{equation}}
\def\ee{\end{equation}}

\def\bea{\begin{eqnarray}}
\def\eea{\end{eqnarray}}

\def\eg{{\it e.g.}}
\def\ie{{\it i.e.}}
\def\etal{{\it et al.}}
\begin{document}
\hspace{1.0cm} \parbox{15.0cm}{

  \baselineskip = 15pt

  \noindent {\bf THE CONSTRAINTS ON POWER SPECTRUM OF RELIC GRAVITATIONAL\\
    WAVES FROM CURRENT OBSERVATIONS OF LARGE-SCALE STRUCTURE OF
    THE UNIVERSE}

  \bigskip \bigskip
  \noindent {\bf B.~Novosyadlyj, S.~Apunevych}

  \bigskip \bigskip

  \baselineskip = 9.5pt

  \noindent {\small {\it Astronomical Observatory of Ivan Franko
      National
      University of Lviv \\
      8 Kyryla i Methodia Str., 79005 Lviv, Ukraine}} \\
  \noindent {\small {\it e-mail:}} {\tt novos@astro.franko.lviv.ua},
  {\tt apus@astro.franko.lviv.ua}

  \baselineskip = 9.5pt \medskip

  \medskip \hrule \medskip

  \noindent Within the framework of cosmological inflationary models,
  relic gravitational waves arise as a natural consequence of the
  quantum nature of primordial space-time metric fluctuations and
  validity of general relativity theory. Therefore the detection of
  cosmological gravitational waves would be the strongest evidence in
  support of basic assumptions of inflation theory.  Although tracing
  gravitational waves by polarization patterns on last scattering
  surface of the cosmic microwave background is only planned for
  forthcoming experiments, some general constraints on the tensor mode
  of metric perturbations (\ie\ gravitational waves) can be
  established right now.

  We present the determination of the amplitude of relic gravitational
  waves power spectrum. Indirect best-fit technique was applied to
  compare observational data and theory predictions.  As observations
  we have used data on large-scale structure of the Universe and
  anisotropy of cosmic microwave background temperature. The
  conventional inflationary model with 11 parameters has been
  investigated, all of them evaluated jointly. This approach gave us a
  possibility to find parameters of power spectrum of gravitational
  waves along with statistical errors.

  \medskip \hrule \medskip }

\vspace{1.0cm}

\renewcommand{\thefootnote}{\ }

\footnotetext{\copyright~B.~Novosyadlyj,~S.~Apunevych~2004}

\renewcommand{\thefootnote}{\arabic{footnote}}

\baselineskip = 11.2pt

\noindent{\small {\bf INTRODUCTION}}
\medskip

\noindent For the last decade due to the progress in observations
the cosmology entered the new stage of its development. This stage was
signalized by the new generation of experiments aimed to the
measurements of anisotropies of cosmic microwave background (CMB).
These were balloon-borne BOOMERanG, MAXIMA, Archeops, ground-based
interferometers DASI, CBI, VSA.  And probably the most important one
is the successor of COBE space mission, Wilkinson Microwave Anisotropy
Probe (WMAP), that published 1-year observation results in 2003. WMAP
has carried out the measurements of CMB over the whole sky with
unprecedented angular resolution and high sensitivity of detectors.
The data and web-links for these experiments are available at the
web-site of Legacy Archive for Microwave Background Data Analysis
(LAMBDA) project\cite{lambda}. CMB explorations were complemented with
extensive measurements of expansion dynamics of the Universe by
distances to Supernovae and large-scale structure surveys.

Advances in quality of experimental data manifestly call for the model
capable to explain the whole set of collected data. Now it's well
understood, that the simplest cosmological models can not
match the observations adequately, as \eg\ standard flat CDM model
with scale-invariant power spectrum of density fluctuations. The
observations advance the complication of model, the number of parameters
increases as well as the number of phenomena encompassed by theory.
Today elaborate inflationary models need about 11 parameters for
the proper description of reality.

Cosmological parameters could be classified in a following way:
\begin{itemize}
\item The parameters related to the background model of the
  homogeneous and isotropic Universe. Evolution of Universe in this
  model is determined by the amount of energy densities of different
  components in ratio to the critical one
  $\Omega_{i}=\rho_{i}/\rho_{cr}$.  These are densities of baryon
  component $\Omega_{b}$, hot dark matter (massive neutrinos)
  $\Omega_{\nu}$, cold dark matter $\Omega_{cdm}$, and density parameter
  for dark energy (\ie\ cosmological constant or quintessence)
  $\Omega_{de}$.
\item The global properties of Universe. According to the Friedmann
  equations sum of density parameters gives unity for the spatially
  flat Universe, or $1-\Omega_{k}$ for curved, so $\Omega_{k}$ is a
  curvature parameter. The Hubble constant $\mathrm{H}_0 = 100h\;
  km/(s\cdot Mpc)$ is regarded as a global parameter too.
\item Parameters associated with the inflation. According to
  inflationary scenario the large-scale structure of Universe is
  assumed to be formed due to growth of primordial matter density
  perturbations because of the gravitational instability.  The
  perturbations have adiabatic nature and originate from quantum
  fluctuations stretched to the cosmological scales during the stage
  of exponential expansion -- inflation.  Perturbations are usually
  described by their power spectra in most general form $P_{s}(k) =
  A_{s} k^{n_{s}}$ for scalar and $P_{t}(k) = A_{t} k^{n_{t}}$ for
  tensor mode of perturbations.  
\item The list of parameters is concluded by some additional
  parameters, like late re-ionization parameter $\tau_c$ and biasing
  parameters to relate the distribution of density to the spatial
  distribution of real astrophysical objects, \eg\ galaxies.
\end{itemize}

This parameter set is required for building the predictions of model
to be compared with the observations for the purpose of finding out
how good the theoretical assumptions match the results of current
observations.  As far as these predictions depend on the parameters
quite nonlinearly, the determination of the best-fit value for one
parameter requests the determination of full set of parameters, by
means of statistics.  Actually that is the main task of this
investigation, sometimes referred as a testing of cosmological models.

In this paper we shall attract attention to the one parameter among
many others, namely the primordial power spectrum of tensor mode of
space-time metric perturbations (relic gravitational waves).  We shall
use the most common kind of cosmological model, and analyze some
particular inflation models.  Sometimes this model is designated as
concordance model, or even standard model.  The purpose of this model
is to give plausible explanation for large-scale structure of the
Universe and its global properties.


\bigskip
\noindent{\small {\bf THE NATURE OF RELIC GWs}}
\medskip

Indeed, gravitational waves are very simple conceptually but proved to
be extremely elusive for detection. Gravitational waves ought to be
emitted by any physical system with changes in quadrupole
distribution of stress-energy density. This follows from equations of
general relativity as a free solution (wave solution) for small linear
space-time metric perturbations on the background metric.

Obviously, that only astrophysical objects could produce gravitational
waves of significant amplitudes. The number of probable astrophysical
sources is under discussion for observational programmes (see for
review \cite{Grishchuk_review}). There is no direct
successful detections by now of any kind of sources. For cosmological
GWs in fact very small chance exist to be detected by any human-made
antenna because of their super-long wavelengths.

Cosmological GWs play the role of most ancient relic in our Universe.
They have their origin in first moments in Universe evolution,
presumably the times of inflation. If we assume general
relativity theory to be valid in those times and quantum zero-point
oscillations to exist, so we come to conclusion of GWs inescapable
existence.  Variable gravitational field in very early Universe
parametrically amplifies quantum oscillations, making up stochastic
gravitational background. The amplitude of these GWs is determined by
energy scales at the moment of emission, so for inflation this
amplitude determines the energy scale of processes that give rise for
inflation.  So the nature of relic gravitational waves is very
fundamental for physics. In fact both scalar mode and tensor modes of
space-time metric perturbations share common origin from single
physical process, quantum oscillations.

The subsequent evolution of metric linear perturbations is completely
described by gauge-invariant formalism for cosmological perturbations.
Relic GWs are represented by tensor mode of space-time metric
perturbations on the isotropic and homogeneous expanding background.
As far as GWs do not interact with rest of medium in expanding
Universe, their amplitude should decrease with time (see
\cite{Durrer_review} for exhaustive review).

Gauge-invariant theory leads to equation for evolution of tensor
perturbations of space-time metrics in vacuum or perfect fluid:
\begin{equation}
  \mathrm{\ddot H}^{(T)}+2 \frac{\dot a}{a} \mathrm{\dot
    H}^{(T)}+(2\mathcal{K}+k^2)\mathrm{H}^{(T)} = 0
\end{equation}
where $\mathrm{H}^{(T)}$ is gauge-invariant amplitude of tensor
perturbations for two polarizations, $a$ -- scale factor,
$\mathcal{K}=(-1,0,1)$ -- curvature index and $k$ is the wavenumber.
The dots denote the derivatives with respect to the conformal time
$\eta$.  The solution of this equation is propagating damped wave.

\bigskip

\bigskip
\noindent{\small {\bf OUTLINE OF METHOD}}
\medskip

As it was stated above both tensor and scalar modes of space-time
metric perturbations have the same origin. The perturbations of scalar
mode connected to the density perturbations so they eventually lead to
the formation of galaxies, clusters and voids, \ie\ large-scale
structure of the Universe. Since tensor perturbations do not lead to
the formation of large-scale structure, cosmological GWs can not be
revealed by present state of observable structure. Both scalar and
tensor perturbations produce the power in CMB angular power spectrum
due to Sachs-Wolfe effect. Of course, because of its specific
polarization properties relic GWs should generate particular
polarization pattern of CMB anisotropies, but the detection of
polarization patterns in CMB is matter of future.

\begin{figure}[htbp]
  \centerline{\includegraphics[width=8.0cm,angle=-90]{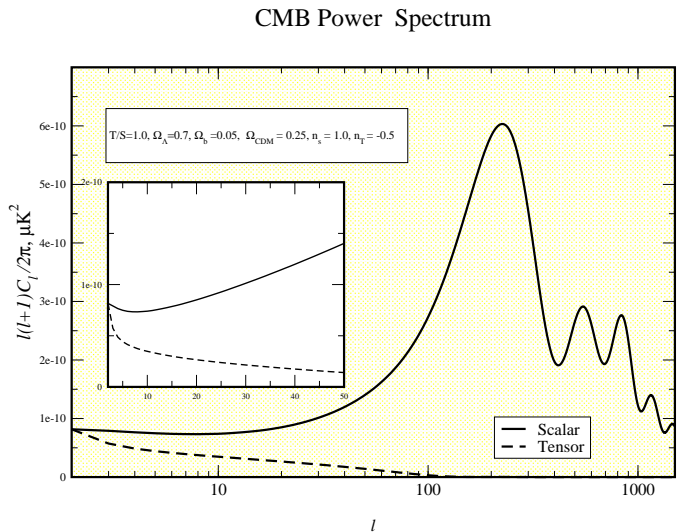}}
  \caption{The scalar and tensor contributions to CMB angular power
    spectrum for particular model.}
  \label{fig:cl_spectrum}
\end{figure}

Therefore we propose the indirect method. As far as CMB power spectrum
consists of contributions from scalar and tensor modes, we can extract
the last one if we know precisely from large-scale structure the
scalar contribution. Also, relic GWs have extra-long wavelengths of
particle horizon size, so the CMB power spectrum from them should
manifest fast decrease with higher multipole moments (smaller scales).
On the contrary, the density perturbations at larger multipoles
produce a feature-rich CMB power spectrum, a series of acoustic peaks.
Thus, the physical properties of tensor mode of cosmological
perturbations of space-time metric can be estimated on the basis of
observational data on: angular power spectrum of CMB combined and
large-scale structure of Universe in wide range of scales -- from
galactic ones to the size of particle horizon, from $10^{-3}$ to
$10^{4}$ Mpc. This way we can constrain total level of intensity of
gravitational waves. In multipoles range of $\ell \sim 10-20$ scalar
and tensor contributions are comparable (Fig.~\ref{fig:cl_spectrum}).
So the precision of result is to be determined by precision of datapoints within
this range. This method is basically a normalization of angular power
spectrum.

\bigskip
\noindent{\small {\bf BUILDING PREDICTIONS}}
\medskip

Unfortunately, the cosmological models are quite complicated and the
most of predictions could not be reduced to analytical functions with
cosmological parameters as arguments. Therefore for the given parameter set
one need to run numerical computations to make predictions. One way is
the ``brute-force'' method, when the numerical packages like CMBfast
\cite{cmbfast} are utilized. Since the testing requires large number
of models to be evaluated, so even quite fast codes need a huge
computational resources to accomplish the task.  Another demerit lies
in so-called ``black box'' uncertainty of numerical calculations, when
the physical processes are hidden inside computations.  So we propose a
semi-analytical approaches to make predictions of CMB power spectrum.

For tensor contribution to CMB power spectrum within low multipoles
range ($2 <\ell < 20$) we have developed an analytical approximation
able to reproduce computed spectrum with high accuracy, for wide range
of parameters values. This approximation takes a form:
\begin{equation}
  \ell (\ell+1) \mathcal{C}_\ell^T = 
  \frac{A(n_t, \Omega_0)}{l+b(n_t,\Omega_0)}\cdot K_\ell(\Omega_{de}) 
  \times \exp(-C(n_t, \Omega_0)\cdot \ell^2 + D(n_t, \Omega_0)\cdot
  \ell)
\end{equation}
where the coefficients $A$, $b$, $C$ and $D$ represent polynomial
functions of $n_t$ and $\Omega_{0} = 1-\Omega_{k}$,
$K_\ell(\Omega_{de})$ is amplification factor for CMB power spectrum
caused by dark energy. Analytical formulae for them presented on our
paper \cite{na2005}.

Semi-analytical approximation for scalar contribution to CMB power
spectrum for the multipoles range $2 <\ell < 20$ was developed in our
previous paper \cite{durrer03} and used here. This
approximation combine the speed of computations and clear unobscured
physical meaning of processes that cause the anisotropies in CMB.  The
angular power spectrum of CMB at higher multipoles have a number of
peaks separated with dips and manifest the damping of average
amplitude to the highest $\ell$. This features are explained by
acoustic oscillations in photon-baryon fluid set up by adiabatic
perturbations on entering the sound horizon. Instead of calculations
of power for each $\ell$ within a wide range, we propose to describe
the whole shape of CMB power spectrum by positions and amplitudes of
the peaks and dips.  The insignificant loss of information will be a
price for the computations speedup. The number of analytical
approximations was developed for heights and positions of first three
peaks and for position of first dip in CMB power spectrum in models
under consideration, see \cite{durrer03}.

\bigskip
\noindent{\small {\bf OBSERVATIONAL DATA}}
\medskip

Of course, the results of testing of the model strongly depend on the
data used. The next important requirement the data should meet is the
statistical independence, \ie\ covariance matrix of datapoints errors
should be diagonal. Here we have used the same set of observational
data, as in papers \cite{durrer01,durrer03} with some changes. The
amplitudes and positions of 1st and 2nd acoustic peaks and position of
first dip have been taken from results of WMAP mission team
\cite{wmap_peaks}, position and height of third peak have been taken
from last recompilation of results of BOOMERanG experiment
\cite{boom2002}.  At large angular scales (lower multipoles range) we
use all datapoints from COBE experiment and WMAP \cite{lambda}, except
dipole.

The CMB data are complemented by large-scale structure data: mass function
and spatial distribution of rich galaxy clusters, temperature function
of X-ray clusters, Ly-$\alpha$ forests of absorption lines in distant
quasars spectra, peculiar velocities of galaxies. These LSS data can
establish the amplitude and shape of spatial power spectrum of matter
density perturbations. Determination of parameters has a problem of
the degeneracy in dependence of observational manifestations upon the
parameters. This problem is partially eliminated when we add
additional measurements to the observational set: determinations of
expansion dynamics by angular distance to Supernovae of Ia type,
independent determinations of Hubble constant, constraints on baryon
content from Big Bang nucleosynthesis theory.

In sum there are 41 values in list of observational datapoints along
with 1$\sigma$ statistical errors of their measurements. We consider all
measurements to be independent and take their probability distribution
as normal distribution. These values and errors are described in
details in cited papers.  Thus if we compare these results with those
from paper \cite{durrer03}, we obtain the answer to the question ``How
improvements in measurements of CMB anisotropies affect the
determination of parameters of observable Universe?''.


\bigskip
\noindent{\small {\bf LIKELIHOOD ANALYSIS}}
\medskip

Lets us have $N$ observational characteristics and we are searching
best-fit values of $n$ cosmological parameters. In other words, we
have the parameter set: $$
\mathbf{\overrightarrow{P}} =
(\Omega_{CDM},\Omega_{de},\Omega_{\nu},N_\nu,\Omega_b,h,A_s,n_s,A_t,n_t,\tau_c)
$$
and the set of observations: $$
\mathbf{\overrightarrow{D}} =
(A_{p_1},\ell_{p_1},A_{p_2},\ell_{p_2},A_{p_3},\ell_{p_3},\ell_{d_1},C_{l_{low}},LSS,NS,h,SN)
$$
To find best-fit values for 11 parameters we are used the
Levenberg-Marquardt algorithm \cite{nr92} for minimization of function
\begin{equation}
  \chi^{2}=\sum_{j=1}^{N}\left({\tilde y_j-y_j
      \over \Delta \tilde y_j} \right)^2~, 
\end{equation}
where $\tilde y_j$ -- observational value for some $j$-th
characteristic, $y_j$ -- its theoretically predicted value, $\Delta
\tilde y_j$ -- statistical uncertainty for measured value.  Since the
number of neutrino species $N_{\nu}$ is a discrete value, so we were
searching for values of 10 parameters at $N_{\nu}$ held fixed at 1, 2
3. As in previous papers, instead of dimensional values of amplitude
of scalar mode power spectra $A_s$, we have used dimensionless value
$\delta_h$, defined as r.m.s. of the fluctuation of matter density on
the scale of present horizon of particle, relate as
$A_s=2\pi^{2}\delta_{h}^{2}(3000{\rm Mpc}/h)^{3+n_s}$.

To find errors for parameters values one needs to carry out
exploration of likelihood function profile $\mathcal{L} \propto
e^{-{1\over 2}\chi^2}$ in parametric space to determine confidential
ranges (marginalizing).  The estimations of confidential ranges for
parameters are mostly based on the direct likelihood function
integration \cite{nr92}, a kind of cumbersome computations.  In papers
\cite{durrer01,durrer03} we have proposed ``economical'' methods for
marginalization procedure to avoid the direct integration. Here we
propose a combined approach: likelihood function profile for parameter
$x_k$ is built using the minimization of $\chi^2$ in subspace of $n-1$
parameters
\begin{equation}
  \mathcal{L}(x_k)=e^{-{1\over 2}[\chi^2(x^{bf}_{i\ne k},x_k)-\chi^2_{min}]}~,
  \label{lhf}
\end{equation} 
where $x^{bf}_{i\ne k}$ -- best-fit values of cosmological parameters
($i=1, 2, \ldots 12$, $i\ne k$), for which $\chi^2$-function has the
minimum at fixed value of parameter $x_k$.  This kind of representation
for $\mathcal{L}(x_k)$ can be obtained from the general integral form.
Numerical experiment proves that results from proposed function
$\mathcal{L}(x_k)$ for confidential levels estimation virtually
coincide with values, obtained by integration.

\bigskip
\noindent{\small {\bf RESULTS}}
\medskip

For historical reasons the common practice is to give a description of
relic gravitational waves by the ratio of amplitudes of tensor $T$ and
scalar $S$ contributions to the quadrupole ($\ell=2$) component of CMB
power spectrum.  However, probability distributions for $A_t$ and
$T/S$ differ, and $\mathcal{L}(A_{t})$ function is closer to the
Gaussian shape than $\mathcal{L}(T/S)$. So we are using
$\mathcal{L}(A_{t})$ for determination of upper bounds and later
recalculate it to $T/S$.  We define the upper bound $A_t^{2\sigma}$ at
confidential level 2$\sigma$ as value for which the square under curve
$\mathcal{L}(A_{t})$ consist 95.4\% of the total square under this
curve from 0 to $\infty$.

\begin{figure}[htbp]
  \centerline{\includegraphics[width=8.0cm]{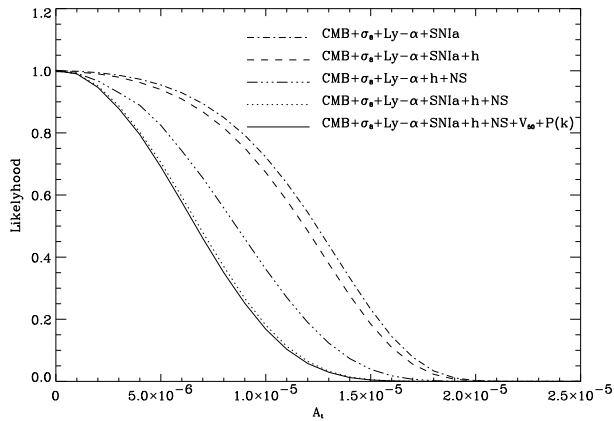}}
  \caption{Likelihood functions $\mathcal{L} (A_{t})=\exp[{-{1\over
        2}\chi(A_t)}]$ for various sets of observational data.}
  \label{fig:results}
\end{figure}

In terms of our technique CMB and LSS data have not sufficient
statistical weight to make possible simultaneous determination of both
amplitude and slope of tensor power spectrum. In other words, if we
would leave both $A_t$ and $n_t$ as free parameters, then $A_t$ could
take arbitrary large value under condition $n_t\rightarrow - \infty$.
If we keep the lower bound for $n_t$ fixed, then upper bound for $A_t$
depends on it. This problem finds its natural explanation by
limitations of indirect method used. We have no data to distinguish
the amplitude and slope separately. Fortunately, the majority of
inflation models relate the $n_t$ with the slope of scalar power
spectrum $n_s$. So we have also analyzed the likelihood functions 
for the same observational data and some generic inflation
models: with flat spectrum of tensor mode ($n_t=0$), natural inflation
with $n_t=n_s-1$ and chaotic inflation with $n_t=0.5(n_s-1)$. The
upper 2$\sigma$ constraints for them are following:
$A_t^{2\sigma}=1.9\cdot10^{-5}$ for the first model and
$A_t^{2\sigma}=1.0\cdot10^{-5}$ for the rest. Corresponding values for
them are $T/S=0.6$ and $0.18$. These three models are more
interesting from the standpoint of manifestations of tensor mode in
the data of observations, so further analysis will include merely
them.  As far as likelihood functions $\mathcal{L}(A_{t})$ for them
are very close it is quite enough to analyze one of them, namely we
take the model with $n_t=0.5(n_s-1)$.

In Fig.~\ref{fig:results} it is shown how observational data influence
the half-width of likelihood function. As we can see, adding the
observational data with constraints on cosmological parameters
themself, like Hubble constant, Big Bang Nucleosynthesis, SNIa
observations and data on large-scale structure, decreases the level of
confidence for the models with high tensor amplitudes.
At the confidential level of 2$\sigma$ (95.4\%) this amplitude cannot exceed
$\sim 20$\% of scalar mode amplitude.  In our previous estimations (see
\cite{durrer03}), based on the data of balloon experiment BOOMERanG,
this limitation was almost four times larger (at 1$\sigma$ C.L. the
ratio was estimated as $T/S=1.7$ for the model with free $n_t$). So we
clearly see the achievements of WMAP with high precision and
covering over all sky.

The results for the whole parameter set are summarized in the
Table~\ref{tab:results}. There the constraints are tabulated for the
case of chaotic inflation. As we can see, the accordance of these
best-fit values with previous determinations \cite{durrer01,durrer03}
is quite satisfactory. The new precise measurements of CMB temperature
anisotropies significantly tightened confidential ranges and lowered
upper constraints for $\Omega_{\nu}$, $\tau_c$ and $T/S$.  Obtained
results agree with determinations of other authors, that used the WMAP
data \cite{wmap2} (those use somewhat different definitions).

\begin{table}[htbp]

  {\footnotesize
    \caption{\small Best-fit values of parameters, lower and 
      upper bounds at  2$\sigma$ (95.4\%) confidential level.}
    \label{tab:results}
    \begin{center}
      \begin{tabular}{lccccccccc}
        \hline\noalign{\smallskip}
        &$\Omega_{de}$ & $\Omega_m$ & $\Omega_{\nu}$ & $\Omega_b$ & $h$ &
        $\delta_h $ & $n_s$ & $T/S$ & $\tau_c$ \\
        \noalign{\smallskip}
        \hline
        \noalign{\smallskip}
        Best-fit & 0.61 & 0.41 & 0   & 0.062 & 0.61 & $4.2\cdot 10^{-5}$ &  0.92 & 0  &  0 \\
        lower bound & 0.52 & 0.31 & 0   & 0.046 & 0.52 & $3.6\cdot 10^{-5}$ & 0.89  & 0  & 0 \\
        upper bound & 0.69 & 0.51 & 0.03& 0.078 & 0.71 & $5.2\cdot 10^{-5}$ & 0.98  & 0.6& 0.15 \\
        \noalign{\smallskip}
        \hline
      \end{tabular}
    \end{center}
  }
\end{table}

\bigskip
\noindent{\small {\bf CONCLUSIONS}}
\medskip

From the standpoint of statistics interpretation of available
observational data on large-scale structure of the Universe and CMB does
not require the presence of considerable amount of relic gravitational
waves.  According to the basic assumptions of early Universe physics
these relic waves have inescapable nature and common origin with
matter density fluctuations seeding the large-scale structure.  Within the
framework of inflation theory the manifestations of gravitational
background can be quite prominent, as long as amplitude of relic GWs
directly connected to energy scales of inflation, \eg\ Grand Unification
Theory energy scale. Thus, the upper bounds on amplitude of GWs appear
to be of the utmost importance for finding constraints on the moment
and energy scales of inflation allowing to discriminate among models of
inflation.

The advances in measurements of CMB in space experiment WMAP
substantially lowered upper bound for amplitude of tensor mode of
perturbations (\ie\ relic gravitational waves) to the level of $T/S\le
0.6$ (95.4\% C.L.) for models with free slope parameter $n_t$. For
models with flat power spectrum of gravitational waves ($n_t=0$), or
some close to that spectra ($n_t\sim 1-n_s$), this limit appears to be
the even lower $T/S\le 0.18$ (95.4\%).



\end{document}